\documentclass[aps,prc,twocolumn,showpacs,amsmath,amssymb,nofootinbib]{revtex4}
\usepackage{epsfig,colordvi}
\usepackage{rotating}
\usepackage{url}
\usepackage{graphicx}
\usepackage{dcolumn}
\usepackage{bm}
\usepackage{epsfig}
\usepackage{amsmath}
\usepackage{makeidx}
\def\be{\begin{equation}}

\def\ee{\end{equation}}
\def\barr{\begin{array}}
\def\earr{\end{array}}

\def\nn8{\nonumber\\[2pt]}
\def\l{\left}
\def\r{\right}
\def\dis{\displaystyle}
\def\ed{\end{document}}

\begin{document}
\title{Deformed shell model study of event rates for WIMP- $^{73}$Ge scattering}

\author{R. Sahu$^{1}$\footnote{rankasahu@gmail.com} and 
V.K.B. Kota$^2$\footnote{vkbkota@prl.res.in}}

\address{$^{1}$ National Institute of Science and Technology,
Palur Hills, Berhampur-761008, Odisha, India}

\address{$^{2}$Physical Research Laboratory, Ahmedabad 380 009, India}

\date{\hfill \today}

\date{\hfill \today}

\begin{abstract}

The event detection rates for the WIMP (a dark matter candidate) are calculated
with $^{73}$Ge as the detector. The calculations are performed within the
deformed shell model (DSM) based on Hartree-Fock states. First the energy levels
and magnetic moment for the  ground state and two low lying positive parity
states for this nucleus are calculated and compared with experiment.  The 
agreement is quite satisfactory. Then the nuclear wave functions are used  to
investigate the elastic and inelastic scattering of  WIMP from $^{73}$Ge. The
nuclear structure factors which are independent of supersymmetric model are also
calculated as a function of WIMP mass. The event rates are calculated for a
given set of  SUSY parameters. The calculation shows that $^{73}$Ge is a good
detector for detecting dark matter.

\end{abstract}


\maketitle

\section{Introduction}

There are overwhelming evidences for the existence of dark matter in the
universe \cite{Jung-96,Debasish}.  However, it has not yet been observed  in
earth-bound experiments nor created at particle colliders. The existence of dark
matter can be inferred from the rotation curves for spiral galaxies,
gravitational lensing in clusters of galaxies, anisotropy in the cosmic
microwave background radiation etc. The standard model of cosmology indicates
that universe hardly contains $\sim 5$\% luminous matter, the remainder is 23\%
non-luminous dark matter and  72\% dark  energy.  The data from the Cosmic
Background Explorer (COBE) \cite{Smoot-1992}  and Supernova Cosmology project
\cite{Gawser-1988} suggest that most of the dark matter is cold. Hot dark matter
which is  moving at relativistic speed can not cluster on galaxy scale. Up to
now, the nature of this matter remains a  mystery. The baryonic cold dark matter
component can be massive compact  halo objects (MACHOs) like neutron stars,
white dwarfs, jupiters etc.  and results of experimental searches suggest that
MACHO fraction should not  exceed ~20\% \cite{Jung-96}. Super symmetric theories
of physics beyond the standard model provide the most promising nonbaryonic
candidates for dark matter. In the simple picture,  the dark matter in the
galactic halo is assumed to be Weakly Interacting Massive Particles (WIMP).
These particles undergo weak interaction and experience the effects of gravity
but do not participate in electromagnetic or strong interactions. The most
appealing WIMP candidate for nonbaryonic cold dark matter is the lightest super
symmetric particle (LSP) which is expected to be a neutral Majorana fermion
travelling with non-relativistic velocities. In recent years, there have been
considerable theoretical and  experimental efforts to detect the WIMP
\cite{Verg-2015}.

Since the WIMP (represented by $\chi$) interacts very weakly with matter,  its
detection is quite difficult.  One possibility to detect WIMP is through  the
recoil of the nucleus in WIMP-nucleus elastic scattering. Popular detector
nuclei are $^{2}$He, $^{19}$F, $^{23}$Na, $^{27}$Al, $^{29}$Si, $^{40}$Ca,
$^{73}$Ge, $^{75}$As, $^{127}$I, $^{134}$Xe and $^{208}$Pb \cite{Divari-2000}.
In WIMP-nucleus scattering, one should consider, in  addition to the scalar
interaction, the spin-spin interaction in which the WIMP couples to the spin of
the nucleus. Exotic WIMPs can lead to large nucleon spin induced cross  sections
which in turn can lead to non-negligible probability for inelastic WIMP-nucleus
scattering \cite{Verg-2015} provided the energy of the excited state is
sufficiently  low as in the $7/2^+$ state of $^{73}$Ge.  Many such nuclei with
excited state very close to the ground state, have recently  been studied, for
example $^{83}$Kr \cite{Verg-2015}, $^{83}$Kr and $^{125}$Te
\cite{Suhonen-2016}, $^{127}$I, $^{129}$Xe, $^{131}$Xe and $^{133}$Cs 
\cite{Suhonen-2009} etc.

There are many experimental efforts to directly detect dark matter by trying to
measure the energy deposited when a WIMP from the Galactic halo scatters from a
nucleus in the detector. Because of the low count rates, the choice of the
detector becomes very important. Searches for spin-dependent interactions
require the use of targets with nonzero spin.  There are many advantages in
choosing lighter target like $^{73}$Ge as a detector. It is better for
detecting light mass dark matter. It is also stable, its natural abundance is
7.75\%. Hence the cost of making the target can be considerably reduced by
choosing this nucleus. Again the low lying excited state $7/2^+$ at 0.0687 MeV
and half-life of 174 ns can  open up new possibilities for spin dependent
inelastic WIMP-nucleus scattering  which can be sizable. The Cryogenic Dark
Matter Search (CDMS) experimental facility \cite{cdms} is  designed to directly
detect the dark matter with $^{73}$Ge as the target nucleus. It has set the most
sensitive limits on the interaction of WIMP with terrestrial materials. The
development of upgrades is underway and will be located at SNOLAB. Another dark
matter experiment is EDELWEISS facility in France \cite{ edelweiss} which uses
high purity  germanium cryogenic bolometers at milikelvin temperatures.  There
are also other experimental attempts using detectors like $^{19}$F,  $^{127}$I,
$^{129,131}$Xe, $^{133}$Cs etc. see for details \cite{Freese-2013}.

The count rate in WIMP-nucleus scattering experiences an annual modulation  as a
result of the earth's orbital motion around the sun.  The Milky Way is assumed
to be surrounded by a halo of non-rotating dark matter. As the solar system 
moves through the Milky Way, it experiences a wind of dark matter. The earth
moves against the  wind for half of the year and experiences the minimum wind
velocity on  December 2. The wind velocity increase as the earth moves in the
opposite direction attaining maximum value on June 2. Annual modulation is a
powerful signature for the existence of dark matter since such modulation is not
exhibited by the background radiation. The DAMA experiments 
\cite{Bernabei-2013} have claimed the observation of the annual modulation  of a
dark matter signal. However this claim has been contested by many other 
experiments. In view of the above, it will be interesting to study the annual
modulation of the event rates theoretically.

The nuclear structure effects are important and should be incorporated in many
of the astro-particle physics problems. The nuclear model should be first tested
regarding its success in describing the properties of nuclei  before applying it
to problems like dark matter detection. The deformed shell model (DSM), based on
Hartree-Fock (HF) deformed intrinsic states with angular momentum projection and
band mixing, is established to be a good model to describe the properties of
nuclei in the mass range A=60-100. See \cite{ks-book} for details regarding DSM
and its applications. The model is found to be quite  successful in describing 
spectroscopic properties including spectroscopy of N=Z odd-odd nuclei with
isospin projection see for example \cite{ga62-2015}, double beta decay
half-lives  \cite{SSK, dbd-2015},  $\mu -e$ conversion in the field of the
nucleus \cite{mu-e-2003} and so on. It will be quite interesting to employ DSM
to calculate the detection rates for the lightest super symmetric particle (a
dark matter candidate) with $^{73}$Ge as the detector. In addition to the
elastic scattering, the inelastic scattering of WIMP from $^{73}$Ge, which has
not been studied up to  now, will also be considered in this article.

Regarding other theoretical works, it may be mentioned that truncated shell
model calculations have been carried out for the studies of event  rates with
the detectors  $^{83}$Kr \cite{Verg-2015},  $^{83}$Kr and $^{125}$Te
\cite{Suhonen-2016}, $^{127}$I, $^{129}$Xe, $^{131}$Xe and $^{133}$Cs
\cite{Suhonen-2009} etc. In these publications, both the  event rates for
elastic and inelastic WIMP-nucleus scattering have been considered. The annual
modulation of signals are also considered. The event rates and annual modulation
signals for $^{23}$Na, $^{71}$Ga, $^{73}$Ge and $^{127}$I cold dark matter
detectors have been  studied using the truncated shell model
\cite{Suhonen-2006}.  However, in this study the authors  restricted to only
elastic scattering. Using the quasi-particle phonon model, Holmlund et
al.\cite{Suhonen-2004} have studied the elastic WIMP detection  rates for
$^{71}$Ga, $^{73}$Ge and $^{127}$I dark matter detectors. There are also
calculations of spin-dependent and  spin-independent  WIMP currents in nuclei
based on chiral effective field theory using large-scale shell model
\cite{currents-1, currents-2, currents-3}. The structure factors for
spin-independent WIMP scattering off xenon has been recently calculated within
the shell model\cite{currents-4}. In these shell model calculations truncation
of the valance space has been done to bring the matrix dimensions to a
manageable size.

Section II gives some details regarding DSM and also about the formulation of
WIMP-nucleus scattering event rates.  The spectroscopic results and also the
results for  elastic and inelastic scattering of WIMP from $^{73}$Ge are
discussed in  Section III. Finally, concluding remarks are drawn in Sect. IV.

\section{Event rates for WIMP-nucleus scattering}

Direct detection of WIMP is most interesting since the earth is washed with
millions of WIMPs every second coming from the galactic halo. The nuclear
recoil in the WIMP-nucleus scattering can be measured with suitable detectors.
The relevant  theory of WIMP-nucleus scattering  are discussed below. In the
expressions for the event rates, the super-symmetric part is  separated from the
nuclear part  so that the role played by the nuclear part becomes apparent. The
nuclear structure calculations are performed within our deformed shell model 
based on Hartree-Fock states.

\subsection{Elastic scattering}

The differential event rate per unit detector mass can be written as 
\cite{Jung-96}

\begin{equation}
dR = N_t\; \phi\; \frac{d\sigma}{d\mid q \mid ^2} f \; d^3 v\; 
d\mid q \mid^2
\label{eqn.1}
\end{equation}

In the above equation, $N_t$ stands for the number of target nuclei per unit
mass which is equal to $1/(Am_p)$, A being the mass number of the nucleus in the
detector and $m_p$ is the proton mass. $\phi$ is the dark matter flux which
is equal to $\rho_0 v/m_\chi$. $\rho_0$ is the  local WIMP density and $m_\chi$
is the WIMP mass. $f$ takes into account the distribution of the WIMP velocity 
relative to the detector (or earth) and also the motion of the sun and earth. 
The distribution is assumed to be Maxwell-Boltzmann type. If we neglect 
the rotation of earth in its own axis, then $v=\mid {\mathbf v}\mid$ is the 
relative
velocity of WIMP with respect to the detector. $q$ represents the momentum 
transfer to the nuclear target. Introducing the dimensionless variable 
$u=q^2b^2/2$ with $b$ as the oscillator length parameter, 
the WIMP-nucleus differential cross section in the laboratory frame is given by
\cite{Suhonen-2016,Suhonen-2009,Suhonen-2006,Suhonen-2004}
\begin{equation}
\frac{d\sigma (u,v)}{du} = \frac{1}{2}\, \sigma_0\,\left(\frac{1}{m_pb}
\right)^2 \,\frac{c^2}{v^2} \,\frac{d\sigma_{A}(u)}{du} \;;
\label{eqn.2}
\end{equation}
with

\be  
\barr{rcl}
\dis\frac{d\sigma_{A}(u)}{du} & = & \l[f_A^0\Omega_0(0)\r]^2 F_{00}(u) +
2f_A^0 f_A^1 \Omega_0(0) \Omega_1(0) F_{01}(u)  \\ 
& &  + \l[f_A^1\Omega_1(0)\r]^2 F_{11}(u) + M^2 \;.
\earr \label{eqn.3}
\ee

In Eq. (\ref{eqn.3}), the first three terms correspond to spin contribution
coming from the axial current and the fourth term stands for the coherent
part coming mainly from the scalar interaction. The coherent part can
be described in terms of the nuclear form factors given as

\be
M^2 = \l(f_S^0\, \l[Z F_Z(u) + N F_N(u) \r] + f_S^1\, \l[Z F_Z(u) - N F_N(u)
\r]\r)^2 \;.
\label{eqn.4}
\ee
If the proton and neutron form factors $F_Z(u)$ and $F_N(u)$ are nearly
equal, then taking $F_Z(u) \approx F_N(u) =F(u)$, we have

\be
M^2 = A^2 \l(f_S^0 - f_S^1\, \frac{A-2Z}{A} \r)^2 \; \mid F(u) \mid^2.
\label{eqn.5}
\ee
Here, $f_A^0$ and $f_A^1$ represent isoscalar and isovector parts of the axial
vector current and similarly  $f_S^0$ and $f_S^1$ represent isoscalar and
isovector parts of the scalar current. These nucleonic current parameters depend
on the specific SUSY model employed. The spin structure functions
$F_{\rho\rho'}(u)$ with $\rho$, $\rho'$ = 0,1 are defined as
\be
\barr{l}
F_{\rho\rho'}(u) = \dis\sum_{\lambda,\kappa}\frac{\Omega_\rho^{(\lambda,\kappa)}(u)
\Omega_{\rho'}^{(\lambda,\kappa)}(u)}{\Omega_\rho(0)\Omega_{\rho'}(0)}\;;\\
\\
\Omega_\rho^{(\lambda,\kappa)}(u) = \sqrt{\frac{4\pi}{2J_i + 1}} \\
                \times \langle J_f \| \dis\sum_{j=1}^A \left[Y_\lambda(\Omega_j)
\otimes \sigma(j)\right]_\kappa j_\lambda(\sqrt{u}\,r_j) 
\omega_\rho(j) \|J_i\rangle
\earr \label{eqn.6}
\ee 
with $\omega_0(j)=1$ and $\omega_1(j)=\tau(j)$; note that $\tau=+1$ for protons
and $-1$ for neutrons. Here $\Omega_j$ is the solid angle for the position
vector of the $j$-th nucleon and $j_\lambda$ is the spherical Bessel function. 
The static spin matrix elements are defined as $\Omega_\rho(0) =
\Omega_\rho^{(0,1)}(0)$. The distribution function $f$ is given by 
\cite{Suhonen-2016}
\begin{equation}
f({\mathbf v}, {\mathbf v}_E) = \frac{1}{(\sqrt{\pi} v_0)^3}
                     e^{-({\mathbf v} + {\mathbf v}_E)^2/v^2_0}
\label{eqn.7}
\end{equation}
In the above equation (\ref{eqn.7}), ${\mathbf v}$ is the velocity of WIMP with 
respect to the  earth containing the detector
and ${\mathbf v}_E$ is the velocity of the earth with respect to the galactic
center given as
\begin{equation}
{\mathbf v}_E = {\mathbf v}_0 + {\mathbf v}_1.
\label{eqn.8}
\end{equation}
In the above ${\mathbf v}_0$ stands for the velocity of the sun with respect to the galactic centre and ${\mathbf v}_1$ is the velocity of earth
with respect to the sun. Assuming that  the polar axis is aligned along the
direction of ${\mathbf v}_1$ and converting the integration variables into 
dimensionless form, the event rate is obtained by integrating
Eq. (\ref{eqn.1}) with respect to $u$, velocity $v$ and angle $\theta$
and can be written as 
\begin{equation}
\langle R \rangle = \int^1_{-1} d\xi  \int^{\psi_{max}}_{\psi_{min}} d\psi 
\int^{u_{max}}_{u_{min}}
G(\psi, \xi) \frac{d\sigma_{A}(u)}{du} du 
\label{eqn.9}
\end{equation}
In the above, $G(\psi, \xi)$ is given by
\begin{equation}
G(\psi, \xi) = \frac{\rho_0}{m_\chi} \frac{\sigma_0}{Am_p} \left(\frac{1}
{m_pb}\right)^2 \frac{c^2}{\sqrt{\pi}v_0} \psi e^{-\lambda^2} e^{-\psi^2}
e^{-2\lambda\psi\xi}
\label{eqn.10}
\end{equation}
Here, $\psi=v/v_0$, $\lambda=v_E/v_0$, $\xi=cos(\theta)$. The values of the
parameters used in the calculation are the following: the WIMP density
$\rho_0 = 0.3 \;Gev/{cm^3}$, $\sigma_0 = 0.77 \times 10^{-38} cm^2$, mass
of proton $m_p = 1.67 \times 10^{-27}$ kg. $m_\chi$ is the WIMP mass. The
velocity of the sun with respect to the galactic centre is taken to be
$v_0 =220$ Km/s and the velocity of the earth relative to the sun is taken as
$v_1=30$ Km/s. The velocity of the earth with respect to the galactic
centre $v_E$ is given by
\begin{equation} 
v_E = \sqrt{v_0^2 + v_1^2 + 2v_0v_1 \sin(\gamma) cos(\alpha)}
\end{equation}
$\alpha$ is the modulation angle which stands for the phase of the earth on 
its orbit around the sun and $\gamma$ is the angle between the normal to
the elliptic and the galactic equator which is taken to be $\simeq 29.8^\circ$.
The value of the oscillator length parameter is taken to be 1.91 fm. In our
earlier work in the calculation of transition matrix  elements for
 $\mu - e$ conversion in $^{72}$Ge \cite{mu-e-2003}, we had taken the 
value of this length parameter as 1.90 fm. Assuming the $A^{1/6}$ dependence,
the value for $^{73}$Ge is taken to be slightly different 1.91 fm.
Writing $\frac{d\sigma_A}{du}$ from Eq. (\ref{eqn.3}) in the form

\be
\barr{rcl}
\dis\frac{d\sigma_{A}(u)}{du} & = &  \l(f^0_A\r)^2 X(1) + 2f^0_Af^1_A X(2)
        +\l(f^1_A\r)^2 X(3)\\
& & + A^2 \l(f_S^0 - f_S^1\, \frac{A-2Z}{A} \r)^2 X(4) \;
\earr \label{eqn.12a}
\ee

\noindent
where $X(1)= [\Omega_0(0)]^2 F_{00}(u)$, 
$X(2)=\Omega_0(0)\Omega_1(0)$ $F_{01}(u)$,
$X(3)=  [\Omega_1(0)]^2 F_{11}(u)$,
$X(4) = [F(u)]^2$,
the event rate per unit mass of the detector can be written as 
\be
\barr{rcl}
\langle R \rangle & = & (f^0_1)^2 D_1 + 2 f^0_Af^1_A D_2 + (f^1_A)^2 D_3+\\
&& A^2 \l(f_S^0 - f_S^1\, \frac{A-2Z}{A} \r)^2 \; \mid F(u) \mid^2 D_4
\earr
\label{eqn.12}
\ee
$D_1$, $D_2$, $D_3$ are the three dimensional integrations of Eq. (\ref{eqn.9})
involving the first three spin
dependent terms of Eq. (\ref{eqn.3}) and $D_4$ is the integration involving
the coherent term. 
\begin{equation}
D_i = \int^1_{-1} d\xi  \int^{\psi_{max}}_{\psi_{min}} d\psi 
\int^{u_{max}}_{u_{min}}
G(\psi, \xi) X(i) du 
\label{eqn.9a}
\end{equation}

The lower and upper limits of integrations given in Eq.(\ref{eqn.9}) and 
(\ref{eqn.9a}) have been worked out by Pirinen et al \cite{Suhonen-2016} and 
they are 
\begin{equation}
\psi_{min} = \frac{c}{v_0} \left(\frac{Am_pQ_{thr}}{2\mu^2_r}\right )^{1/2}
\label{eqn.13}
\end{equation}
\begin{equation}
\psi_{max} = -\lambda\xi + \sqrt{\lambda^2\xi^2+\frac{v_{esc}^2}{v_0^2} -1
- \frac{v^2_1}{v^2_0}-\frac{2 v_1}{v_0} sin(\gamma) cos(\alpha)}
\label{eqn.14}
\end{equation}
Taking the escape velocity $v_{esc}$ from our galaxy to be 625 km/s, the
value of $v_{esc}^2/v_0^2 -1- v^2_1/v^2_0$ appearing in 
Eq. (\ref{eqn.14}) is $7.0525$. Similarly taking $\gamma=29.8^\circ$, the value
of $(2 v_1/v_0) sin(\gamma)$ is $0.135$. The values of $u_{min}$ and
$u_{max}$ are
\begin{equation}
u_{min}=Am_pQ_{thr}b^2
\label{eqn.15}
\end{equation}
\begin{equation}
u_{max}=2(\psi\mu_rbv_0/c)^2
\label{eqn.16}
\end{equation}
In the above equations, $Q_{thr}$ is the detector threshold energy and $\mu_r$
is the reduced mass of the WIMP-nucleus system.

\subsection{Inelastic scattering}

In the inelastic scattering the entrance channel and exit channel are
different and hence the coherent part does not contribute to the event rate. 
Hence the inelastic event rate per unit mass of the detector can be obtained
from Eq. (\ref{eqn.12a})
\begin{equation}
\langle R \rangle_{in} = (f^0_1)^2 E_1 + 2 f^0_Af^1_A E_2 + (f^1_A)^2 E_3
\label{eqn.19}
\end{equation}
$E_1$, $E_2$ and $E_3$ are the three dimensional integrations
\begin{equation}
E_i = \int^1_{-1} d\xi  \int^{\psi_{max}}_{\psi_{min}} d\psi 
\int^{u_{max}}_{u_{min}}
G(\psi, \xi) X(i) du 
\label{eqn.20}
\end{equation}
The integration limits for $E_1$, $E_2$ and $E_3$ are \cite{Suhonen-2016, 
Suhonen-2009} 
\begin{equation}
u_{min} = \frac{1}{2}b^2\mu_r^2\frac{v^2_0}{c^2}\psi^2
\left[ 1 - \sqrt{1-\frac{\Gamma}{\psi^2}} \right ]^2
\label{eqn.21}
\end{equation}

\begin{equation}
u_{max} = \frac{1}{2}b^2\mu_r^2\frac{v^2_0}{c^2}\psi^2
\left[ 1 + \sqrt{1-\frac{\Gamma}{\psi^2}} \right ]^2
\label{22}
\end{equation}
where 
\begin{equation}
\Gamma = \frac{2 E^*}{\mu_rc^2} \frac{c^2}{v_0^2}
\label{23}
\end{equation}
$E^*$ being the energy of the excited state. $\psi_{max}$ is same as in the
elastic case and the lower limit  $\psi_{min} = \sqrt{\Gamma}$. The values of
the parameters like $\rho_0$, $\sigma_0$ etc. are same as in the elastic case.
\subsection{Deformed shell model}

In the calculation of the event rate (both for elastic and inelastic
scattering), the nuclear part has been separated from the super-symmetric
part in the formulation. The nuclear part enters in the calculation
of the nuclear structure factors $D_1$, $D_2$, $D_3$ and $D_4$ in the
elastic scattering and the corresponding factors $E_1$, $E_2$ and
$E_3$ in inelastic scattering. The evaluation of these nuclear structure
factors in turn depends on  spin structure 
functions  and the form factors. We have used our deformed shell model
for the evaluation of these quantities. The details of this model have 
been described in many of our earlier publications, see for example
\cite{ks-book}. In this model, 
for a given nucleus, starting with a model space consisting of
the given set of single particle (sp) orbitals and effective two-body
Hamiltonian (TBME + spe), the lowest energy intrinsic states are obtained by
solving the Hartree-Fock (HF) single particle equation self-consistently.
 We assume axial symmetry.
Excited intrinsic configurations are obtained by making particle-hole
excitations over the lowest intrinsic state.  These intrinsic states
$\chi_K(\eta)$ do not have definite angular momenta.  Hence states of good 
angular momentum projected from an intrinsic state $\chi_K(\eta)$ can be 
written in the form
\begin{equation}
\psi^J_{MK}(\eta) = \frac{2J+1}{8\pi^2\sqrt{N_{JK}}}\int d\Omega D^{J^*}_{MK}(\Omega)R(\Omega)| \chi_K(\eta) \rangle 
\label{eqn.24}
\end{equation}
where $N_{JK}$ is the normalization constant given by
\begin{equation}
N_{JK} = \frac{2J+1}{2} \int^\pi_0 d\beta \sin \beta d^J_{KK}(\beta)\langle \chi_K(\eta)|e^{-i\beta J_y}|\chi_K(\eta) \rangle  
\label{eqn.25}
\end{equation}
In Eq. (\ref{eqn.24}), $\Omega$ represents the Euler angles ($\alpha$, $\beta$,
$\gamma$), $R(\Omega)$ which is equal to exp($-i\alpha J_z$)exp($-i\beta
J_y$)exp( $-i\gamma J_z$) represents the general rotation operator.  The good
angular momentum states projected from different intrinsic states are not in
general orthogonal to each other.  Hence they are orthonormalized and then 
band mixing calculations are performed.  DSM is well established to be a 
successful model for transitional nuclei (with A=60-90)
\cite{ks-book,ga62-2015,SSK,KS1,KS2,KS3}. 

In the evaluation of the  reduced matrix element appearing in Eq.
(\ref{eqn.6}) in DSM, we need the sp matrix elements of
the operator of the form $t^{(l,s)J}_\nu$ and these are given by,
\be
\barr{l}
\langle n_il_ij_i\| \hat{t}^{(l,s)J}\| n_kl_kj_k \rangle = \\
\\
\dis\sqrt{(2j_k+1)(2j_i+1)(2J+1)(s+1)(s+2)} \\
\\
\left\{\begin{array}{ccc}
l_i & 1/2 & j_i\\
l_k & 1/2 & j_k\\
l   & s   & J
\end{array}\right\} \;
\langle l_i \| \sqrt{4\pi}Y^l \|l_k \rangle \;\langle n_il_i \|j_l(kr)\| 
n_kl_k \rangle \;.
\earr \label{eqn.26}
\ee
In the above equation, $\{--\}$ is the nine-$j$ symbol. \\

\section{Results and discussions}

\begin{figure} 
\includegraphics[width=8.5cm]{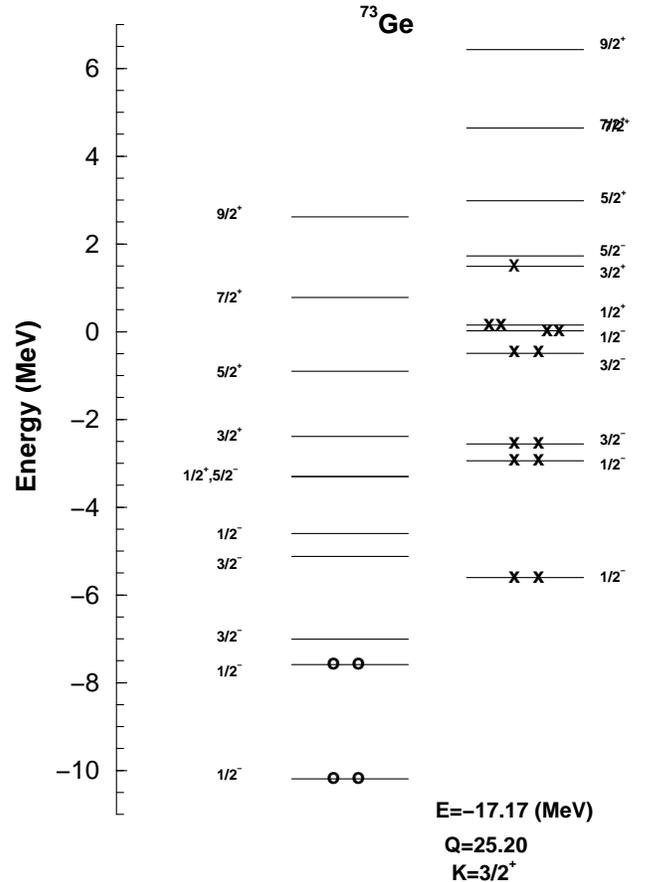} 
\caption{HF single-particle spectra for $^{73}$Ge corresponding to the lowest
prolate configuration. In the figure circles represent protons and
crosses represent neutrons. The HF energy (E) in MeV, mass quadrupole moment (Q)
in units of the square of the oscillator length parameter and the total 
azimuthal quantum number K are given in the figure.}
\label{ge73hf}
\end{figure}

\begin{figure}
\includegraphics[width=6cm]{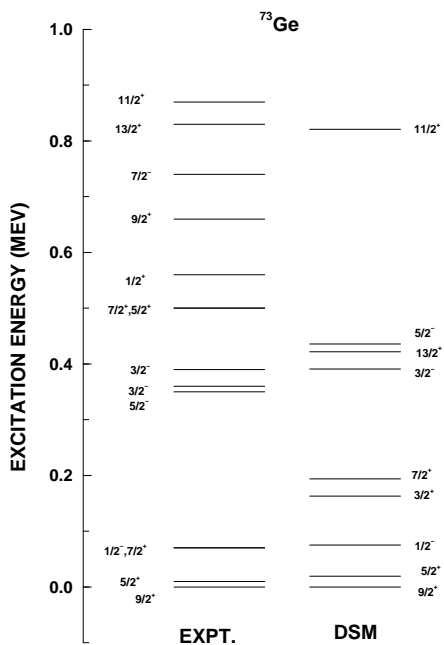}
\caption{ Comparison of deformed shell model results with experimental data 
for the low-lying levels. The experimental values are taken from
\cite{nndc}
}  
\label{ge73sp}
\end{figure}

\begin{figure}
\includegraphics[width=6cm]{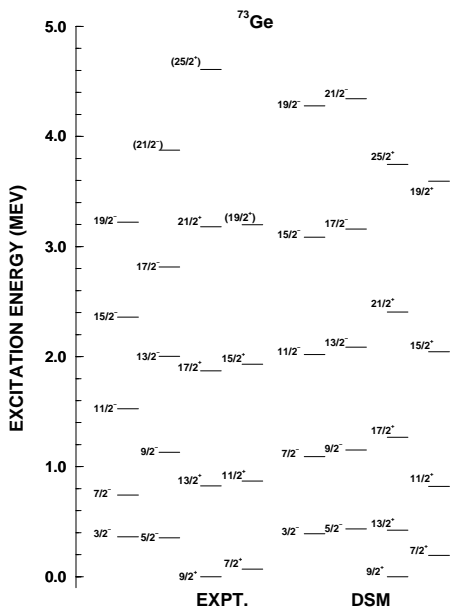}
\caption{ Comparison of deformed shell model results with experimental data 
for the collective bands recently observed for this nucleus 
\cite{jjsun-2015}
}  
\label{ge73hsp}
\end{figure}

Before carrying out the calculation for event rates in WIMP-nucleus scattering,
we first check the goodness of the nuclear wave functions by computing the 
energy level spectrum and magnetic moments and compare them with experimentally
measured quantities. Good agreement with experiment will give us confidence in
our results for event rates. In  DSM calculations, $^{56}$Ni is taken as the
inert core   with  the spherical orbits $1p_{3/2}$, $0f_{5/2}$, $1p_{1/2}$ and
$0g_{9/2}$  forming the basis space. Modified Kuo interaction with single
particle energies 0.0, 0.78, 1.08 and 4.90 MeV has been used in the calculation.
This effective interaction has been used in many of our calculations and has
been found to be quite successful in describing most of the important features
of nuclei in this region. The HF single particle spectrum obtained by solving
the HF equation self-consistently is shown in Fig. \ref{ge73hf}. The four
protons of this nucleus occupy the lowest two $k=1/2^-$ orbits. On the other
hand ten neutrons occupy $pf$ orbits with three remaining neutrons in the
$g_{9/2}$ orbit. As described earlier, we have generated three intrinsic states
of positive parity and three intrinsic states of negative parity by
particle-hole excitation. Good angular momentum states are projected from each
of these intrinsic states and then a band mixing calculation is performed. The
low-lying levels obtained in the band mixing calculation are compared with
experiment as shown in Fig. \ref{ge73sp} up to an excitation energy of 1 MeV.
The data are taken from ref. \cite{nndc, jjsun-2015}. As can be seen, the
agreement with experiment is quite  satisfactory. The ground state and next two
excited states are quite nicely reproduced. However, there are more levels in
experiment than in our calculation. This problem can be solved by taking more
intrinsic states in our calculation. Since we are concerned with the calculation
of event rate for dark matter - nucleus scattering, we are interested only in
the ground state for elastic scattering and ground state and an excited state in
the inelastic scattering, the intrinsic states taken are expected to be
sufficient. The collective bands recently observed by Sun et al 
\cite{jjsun-2015} are compared with our DSM calculated results in  Fig.
\ref{ge73hsp}. The agreement is reasonably good. More number of intrinsic
states  would have improved the results considerably. But our aim here is not to
concentrate in detailed spectroscopy and hence taking more intrinsic states to
study alignment effects in the collective bands are not necessary.

Since spin contributions play an important role in the calculation of event
rates, the calculation of magnetic moment is first carried out. For better
physical insight, it is decomposed into orbital and spin parts. The results for
the contribution of protons and neutrons to the orbital and spin parts for the 
first three positive parity levels $9/2^+$, $7/2^+$ and $5/2^+$ are given in
Tab. \ref{tab-1}. The calculated magnetic moments are also compared with
experiment. In the calculation, bare values of g-factors has been used and no
quenching has been done. The DSM value for the magnetic moment of the ground
state  $9/2^+$ is $-0.811$ $\mu_N$ which agrees quite well with the experimental
value $-0.879$ $\mu_N$. The calculated magnetic moment of the lowest $5/2^+$
also agrees quite well with experiment. However no experimental results are
available for the first $7^+$ level. Holmlund et al. \cite{Suhonen-2004} have
calculated the magnetic moment and also the decomposed orbital and spin parts
for the ground state of this nucleus and  compared their results with different
theoretical calculations. Our values agree quite well with the shell model
results cited in the above publication.

With the success of the DSM model in explaining the low lying energy levels
and the magnetic moments of the lowest three positive parity levels, we then
proceed to calculate the event rates for the elastic and inelastic scattering
with confidence.

\subsection{Results for elastic scattering}

\begin{figure*}
\includegraphics[width=6.5cm]{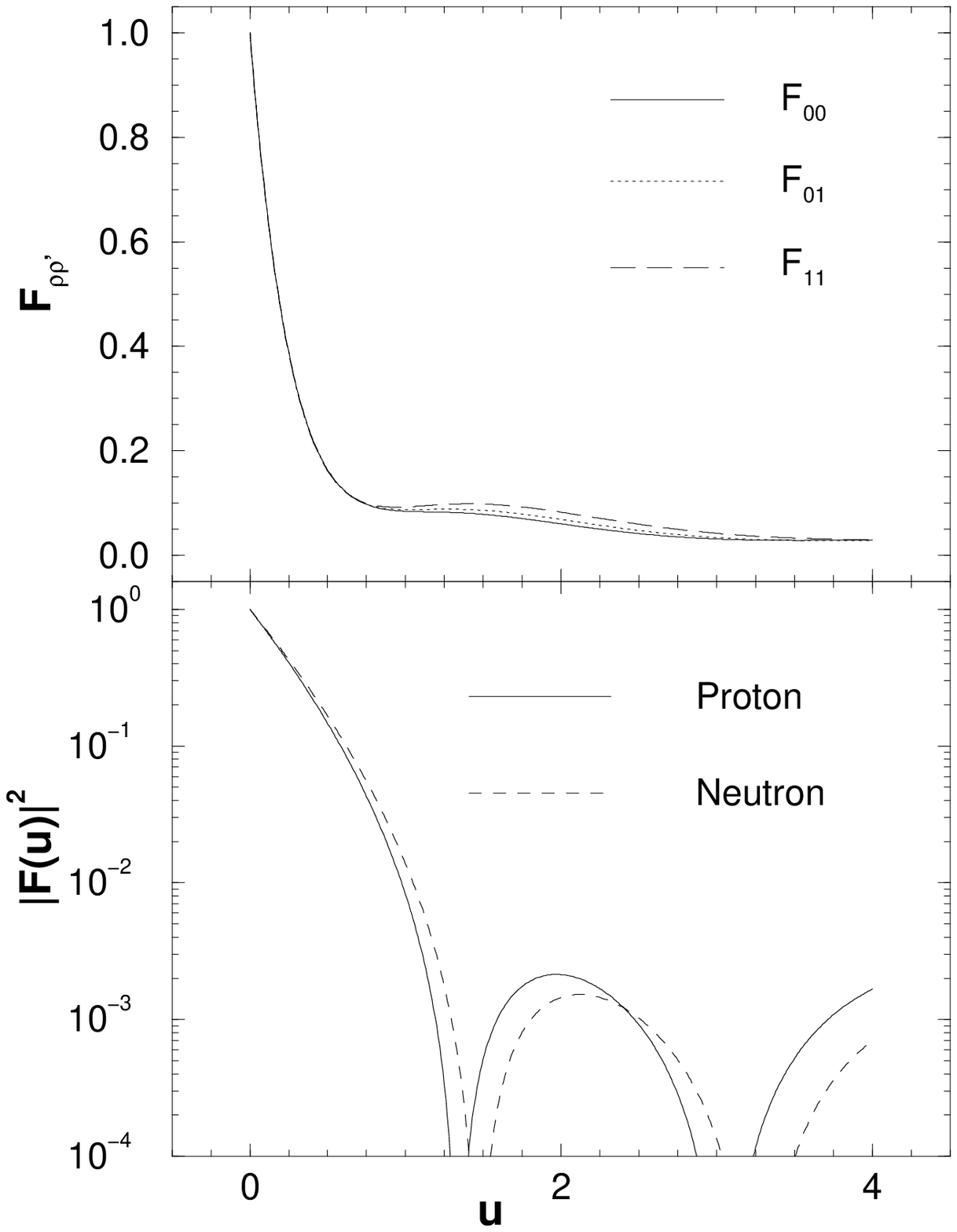}
\caption{Spin structure functions and squared proton and neutron form factor
for $^{73}$Ge.}
\label{ssf-el}
\end{figure*}

Using the DSM wave function, the spin structure functions for the elastic 
channel are calculated  using the formula given  in Eq. (\ref{eqn.6}) for the
ground state $9/2^+$. The results are  plotted in Fig. \ref{ssf-el}.   Similarly
the proton and neutron form factors are calculated using  Eq. (\ref{eqn.6}) and
plotted in Fig. (\ref{ssf-el}). As can be seen, the form factors for  protons
and neutrons are almost same. Hence the approximation made in deriving the Eq.
(\ref{eqn.5})  is correct. We find that $F_{00}$, $F_{01}$ and $F_{11}$ have
almost equal values. The spin structure functions as well as the form factor
approach zero for $u > 1$. Hence the main contribution to the event rate comes
from the spin structure functions and form factor with $u < 1$. The plot of the
spin structure functions is almost similar to results  obtained in ref.
\cite{Suhonen-2004} within their quasi-particle-pairing model. DSM gives the
values of static spin matrix elements $\Omega_0$ and $\Omega_1$  to be 0.798 and
-0.803. These values agree quite nicely with the results obtained by Kortelainen
et al. \cite{Suhonen-2006}. Using shell model Ressel et al. \cite{Ressel-1993}
have calculated $S_{\rho\rho'}$, that are related to the spin structure
functions defined above. Their values compare well with the DSM results.

The coefficients $D_1$, $D_2$, $D_3$ and $D_4$ defined in Eq. (\ref{eqn.9a})
depend only on the nuclear wave functions and they are evaluated within our DSM
model. Since $\Omega_1(0)$ is negative, the coefficient $D_2$ is also negative.
We plot in Fig. \ref{el-dn}
$D_1$, $-D_2$, $D_3$ and $D_4$ as a function of the WIMP mass for three
values of the detector threshold energy $Q_{thr}$= 0, 5, 10 keV. For  $Q_{thr}$
= 0, all the four graphs peak at WIMP mass $\sim$35 GeV. For higher values of 
$Q_{thr}$, the peaks shift towards higher WIMP mass. The height of the peaks for
$D_1$, $-D_2$ and $D_3$ are  smaller than the values reported for $^{83}$Kr
\cite{Suhonen-2016}. However, $D_4$ is almost similar to $^{83}$Kr.  The
thickness of the graphs represents annual modulation. The modulation  signal is
large for WIMP mass below $\sim$70 GeV. However, modulation effect tapers off at
higher values of the WIMP mass. The magnitude of the modulation at WIMP mass
$\sim$35 GeV is about 3.3\% for the spin dependent channels and about 4\% for
the spin independent case. Thus we predict much smaller modulation compared to
$^{83}$Kr where it is $\sim$ 10\%.  At WIMP mass 150 GeV, the modulation effect
for the spin independent channel $D_4$ changes sign.  However, no such reversal
is found for spin dependent channels. As has been  discussed earlier, annual
modulation is a powerful signature for the existence of dark matter since such
modulations are not exhibited by the back ground. The DAMA experiments
\cite{Bernabei-2013}  had claimed the observation of  a positive  dark matter
signal through the observation of annual modulation over 13 years of operation.
However these results are in apparent contradiction with  the null results from
other experiments like CoGent  \cite{Aalseth-2013},  CDMS  \cite{cdms}, XENON
\cite{xenon} experiments. However, the  CRESST experiment \cite{cresst} claims
to have observed the modulation.  Hopefully in the near future, we will get an
unanimous answer regarding the observation of modulation.

\begin{figure}
\includegraphics[width=6.5cm]{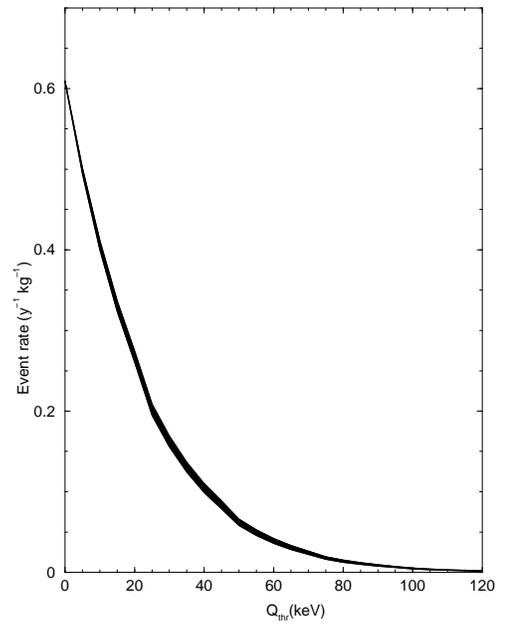}
\caption{The even rate in units of $yr^{-1} kg^{-1}$ as a function of
detector threshold $Q_{thr}$. The thickness of the curve
 represents the annual modulation. 
}
\label{rate-el}
\end{figure}

Using the values of SUSY parameters $f^0_A=3.55e-2$, $F^1_A=5.31e-2$,
$f^0_S=8.02e-4$ and $f^1_S=-0.15\times f^0_S$, we have calculated the  event
rate for the detection WIMP corresponding to the WIMP mass 110 GeV. The results
are given in Fig.  \ref{rate-el}. The thickness of the graph represents
annual modulation.  The event rate changes drastically depending
on the choice of the SUSY parameters \cite{Suhonen-2006}.  This is because, the
expression for event rate can be split into two parts. The spin dependent term
is represented by the coefficients $D_{1,2,3}$ and the coherent term represented
by $D_{4}$. The different parametrizations weigh  these channels differently and
hence the different choice of the parameter set gives different results.

\begin{table} 
\caption{The calculated magnetic moments and their decomposition into orbital
and spin parts for $^{73}$Ge. The experimental values
are given within the parentheses. Bare gyro magnetic ratios have been used
in the calculation. The experimental data are from ref. \cite{nndc}.}
\begin{tabular}{cccccc}
\hline
$J$   & $<l_p>$ & $<S_p>$ & $<l_n>$ & $<S_n>$ & $\mu$ (n.m.) \\
\hline
$9/2^+$   & 0.581   & -0.001 &  3.558  &  0.362  & -0.811\\
          &         &         &         &         & (-0.879) \\
$7/2^+$   & 0.691   & -0.009  & 2.590   &  0.228  & -0.232\\
$5/2^+$   & 0.036   &  0.008  & 2.189   &  0.265  & -0.929\\
          &         &         &         &         & (-1.080) \\
\hline
\end{tabular}
\label{tab-1}
\end{table}

\subsection{Results for inelastic scattering}

In $^{73}$Ge, the first excited positive parity state is $5/2^+$ and scattering
to this state from the ground state is not allowed because of mainly the $M1$
type of the interaction. The next excited positive parity state is $7/2^+$ state
at an excitation energy of 68.7 keV \cite{jjsun-2015} and scattering to this
state is allowed. Unlike the ground state and the first excited positive parity
state for which the magnetic moments have been known, the magnetic moment for
the $7/2^+$ has not been measured, see Tab. \ref{tab-1}. The calculated value is
$-0.232 \; \mu_N$  which should be nearer to the actual value since the magnetic
moment of the other two lower positive parity states agree so well with
experiment.

\begin{figure}
\includegraphics[width=6cm]{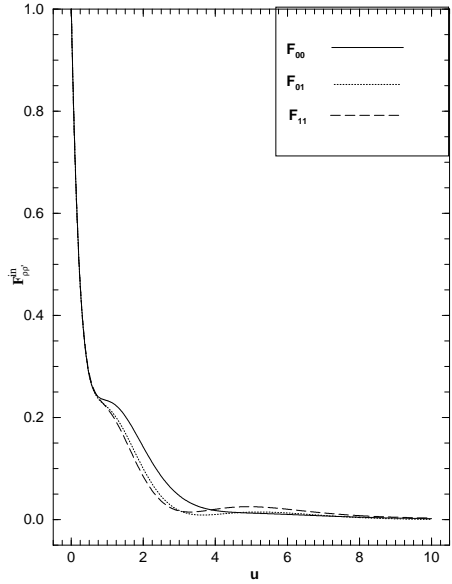}
\caption{Spin structure function in the inelastic channel}
\label{ssf-in}
\end{figure}

The static spin matrix elements for the inelastic scattering to the  $7/2^+$
state are $\Omega_0 = -0.167$ and $\Omega_1=0.142$. These values are about $6-7$
times smaller than the corresponding values in the elastic scattering  case. 
However, these values are  about $4$ times larger than the values quoted  for 
$^{83}$Kr \cite{Suhonen-2016} obtained within the full shell model with $jj44b$
effective interaction.  Our values are almost as large as in $^{125}$Te 
\cite{Suhonen-2016} and
therefore the inelastic  event rate should be competitive. The inelastic spin
structure functions are  presented in Fig. \ref{ssf-in}. The structure functions
almost overlap with each other except for a small window lying between $0.7 \leq
u \leq 4$. In $^{83}$Kr, the three structure functions deviate from each other
over a longer domain.

\begin{figure}
\includegraphics[width=8.5cm]{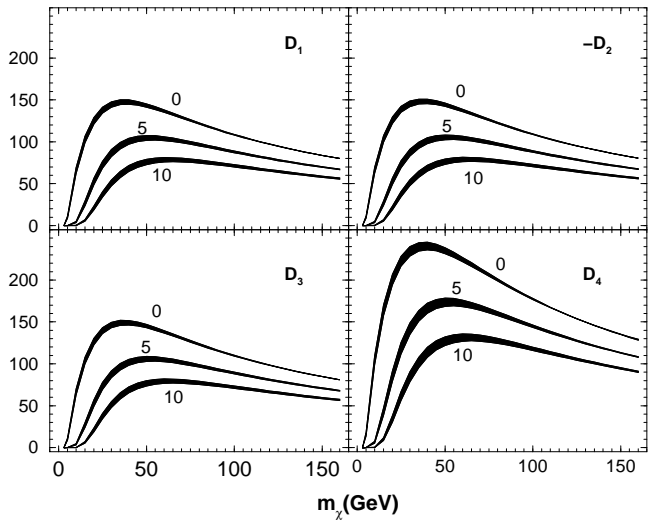}
\caption{Nuclear structure coefficients $D_n$ of Eq. \ref{eqn.9a}
plotted as a function
of the WIMP mass. The graphs are plotted for three values of the 
detector threshold $Q_{thr}$ namely $Q_{thr}=0,5,10$ keV. The close lying graphs
for each value of $Q_{thr}$ represent the annual modulation. 
}
\label{el-dn}
i\end{figure}

The nuclear structure coefficients $E_1$, $E_2$ and $E_3$ defined in Eq.
(\ref{eqn.20}) are calculated and plotted in Fig. \ref{el-dn}. The peaks occur
at around $m_\chi=45$ GeV, almost similar to the elastic case. The peak values
are  almost 13 times larger than the values reported for $^{83}$Kr, but about a
factor of two smaller than for $^{125}$Te \cite{Suhonen-2016}. The modulation is
about 4\% almost similar to the elastic case. But for  $^{83}$Kr and $^{125}$Te,
Pirinen et al. \cite{Suhonen-2016} found the modulation effect to be much larger
in the inelastic case than in elastic case. The nuclear structure coefficients
do not depend on the detector threshold energy. Hence the event rate can be
calculated by reading the values of  $E_i$ from the graph and using the SUSY
parameters. Because of the large values of $E_i$, the inelastic scattering of
WIMP from $^{73}$Ge is  a potential candidate for dark matter detection. \\

\begin{figure}
\includegraphics[width=6cm]{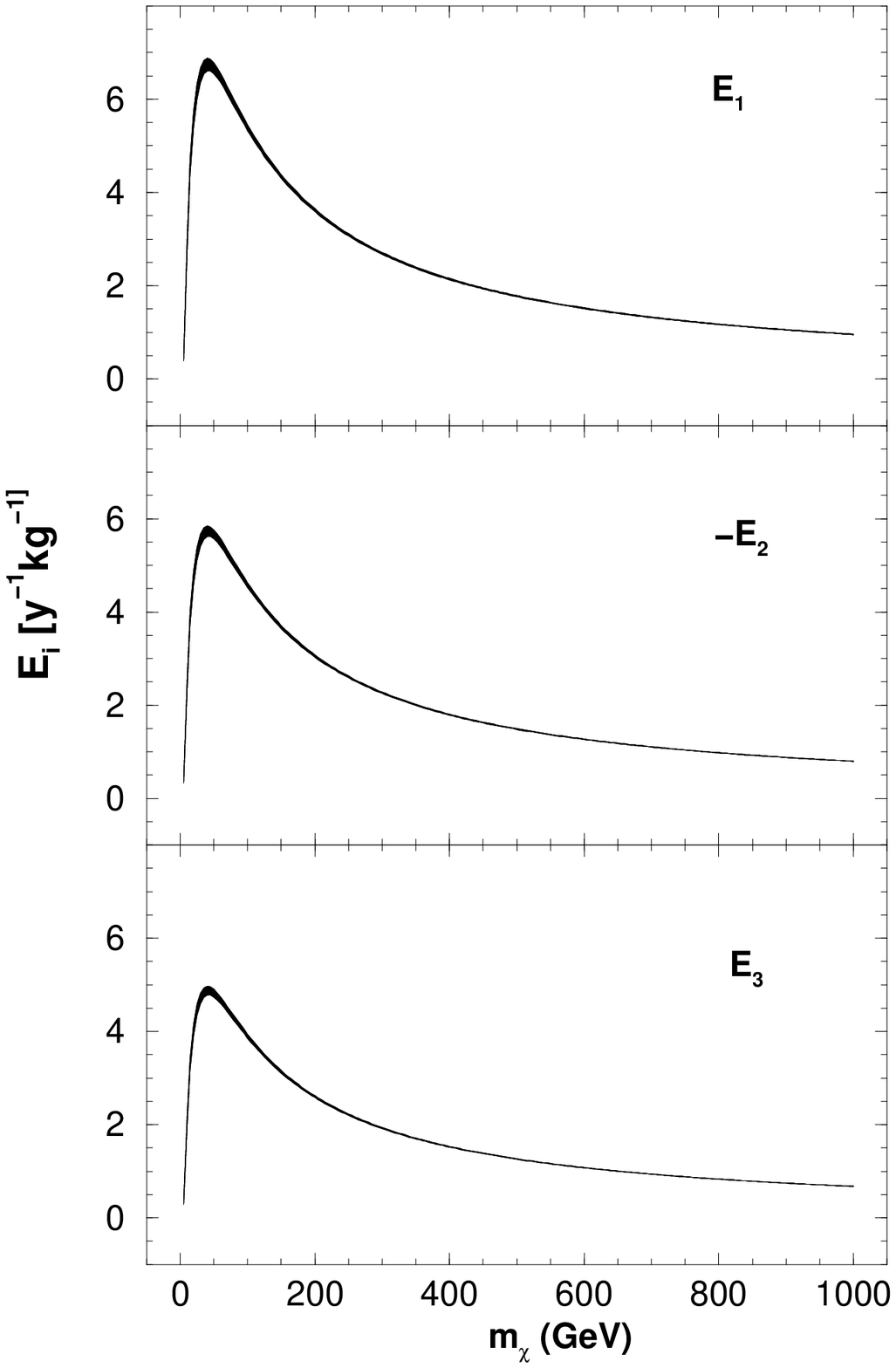}
\caption{Nuclear structure coefficients $E_n$ in the inelastic channel.
The thickness of the graphs represent annual modulation
}
\label{ee-in}
\end{figure}

\section{Conclusions}

The deformed shell model based on HF states is used to study the event rates for
elastic and inelastic scattering of WIMP from $^{73}$Ge. First energy spectrum
and magnetic moments are calculated to test the goodness of the nuclear wave
functions. After ensuring the good agreement with experiment, we studied the
event rate. The nuclear structure  coefficients  and the static spin matrix
elements suggest that $^{73}$Ge is a good detector for dark matter detection. We
have also calculated the annual modulation rates for both elastic and inelastic
scattering which is about 3-4\% at the peaks. The inelastic scattering of WIMP
from $^{73}$Ge has been calculated for the first time in this article. The
nuclear structure coefficients for this nucleus in the inelastic scattering is
much larger than  $^{83}$Kr and are almost as large as $^{125}$Te. Since
$^{73}$Ge is stable and natural abundance is relatively large,  it will be worth
while to experimentally study inelastic scattering from this nucleus.
The detection of the inelastic event rate will provide information regarding
the spin dependent nature of WIMP-nucleus interaction.

\acknowledgements

We are thankful to J.D. Vergados and J. Suhonen for useful correspondence and 
encouragements. Thanks are also due to T.S. Kosmas for discussions in the 
initial stages of this work.  R. Sahu is thankful to SERB of Department of 
Science and Technology (Government of India) for financial support.

\end{document}